\documentclass[final]{article}
\pdfoutput=1

\usepackage{nips_2016}

\newtheorem{theorem}{Theorem}

\usepackage[utf8]{inputenc} 
\usepackage[T1]{fontenc}    
\usepackage{hyperref}       
\usepackage{url}            
\usepackage{booktabs}       
\usepackage{amsfonts}       
\usepackage{nicefrac}       
\usepackage{microtype}      
\usepackage{graphicx}

\title{Lethean Attack: An Online Data Poisoning Technique}

\author{
  Eyal Perry \\
  MIT Media Lab\\
  Massachusetts Institute of Technology\\
  \texttt{eyalp@mit.edu} \\
}

\begin{document}

\maketitle

\begin{abstract}
  Data poisoning is an adversarial scenario where an attacker feeds a specially crafted sequence of samples to an online model in order to subvert learning. We introduce Lethean\footnote{\textit{Lethe}: a river in Hades whose waters cause drinkers to forget their past.} Attack, a novel data poisoning technique which induces catastrophic forgetting on an online model. We apply the attack in the context of Test-Time Training, a modern online learning framework aimed for generalization under distribution shifts. We present the theoretical rationale and empirically compare against other sample sequences which naturally induce forgetting. Our results demonstrate that using lethean attacks, an adversary could revert a test-time training model back to coin-flip accuracy performance using a short sample sequence.
\end{abstract}

\section{Introduction}

Modern machine learning models, especially deep neural networks, have seemingly achieved human performance in problems such as image classification \citep{krizhevsky2012imagenet}, speech recognition \citep{hinton2012deep}, natural language understanding \citep{devlin2018bert} and others. One of the major underlying assumptions behind neural networks is for the training data and test instances be drawn from the same distribution. However, under minor differences in the distribution state-of-the-art models often crumble \citep{sun2019testtime}. Methods such as domain adaption \citep{DBLP:journals/corr/TzengHSD17, ganin2015domainadversarial} and adversarial robustness \citep{aleks2017deep} have tackled shifts in the distribution by either assuming access to data from the new distribution at training time, or assuming a structure of the perturbations. 

Online learning \citep{shalev2012online} is a domain of machine learning, where a learner is given a sequence of instances from a distribution $\mathcal{X}$. For each instance $x_t$, the learner makes a prediction and updates the model. In the classical online learning setting, there exists an oracle that gives the ground truth $y_t$ for each time step $t$. One of the advantages of an online learning paradigm is the ability to adapt a model under distribution shifts. A well-researched pitfall of online neural networks is catastrophic forgetting \citep{mccloskey1989catastrophic, ratcliff1990connectionist}, the tendency of a model forget previously learned information upon the arrival of new instances.

\cite{sun2019testtime} recently proposed a novel online approach named Test-Time Training to achieve better generalization under distribution shifts, without prior knowledge. In summary, test-time training adapts model parameters at testing time, using a self-supervised task. Let's consider the supervised setting, e.g. image classification. Alongside the main task, test-time training requires another self-supervised auxiliary task, e.g. predicting image rotation by an angle (0$^\circ$, 90$^\circ$, 180$^\circ$, 270$^\circ$). The model predicts two outputs (e.g. image class and rotation) and importantly, a significant percentage of the model parameters are shared across the two tasks. \cite{hendrycks2019using} showed that joint-training with a self-supervised task increases robustness of the model. Yet, \cite{sun2019testtime} take one step further than joint-training. At testing time, the new instance $x$ is first handled by the self-supervised task. Then, the model parameters are updated according to the sub-task loss. Finally, the adapted model is run on $x$ with the main task to get a prediction.

\cite{sun2019testtime} presented promising results. By applying the test-time training paradigm to image classification problems, their neural network successfully adapted to various types and levels of corruptions in both CIFAR-10 \citep{krizhevsky2009learning} and ImageNet \citep{russakovsky2015imagenet} and surpassed state-of-the-art models by a large margin. The most significant improvement occurs in an online setting, where at each time step $t$ the model is adapted and saved for future test instances. It is important to note the difference from classical online learning. While both frameworks perform models updates at each time step, test-time training does not assume or need any oracle. It adapts to new instances as they arrive regardless of labels, simply according to their distribution.

By merging the training and testing phases, test-time training has become exposed to an adversarial scenario named data poisoning \citep{zhang2019online, wang2018data}. In online data poisoning, the attacker feeds the model a sequence of specially crafted samples in order to manipulate learning. Moreover, \cite{goodfellow2013empirical} demonstrated that multi-task models are especially prone to a catastrophic forgetting. In this work, we propose a practical methodology for test-time training data poisoning, lethean attack, which induces catastrophic forgetting on a model. First, we describe the attack method and the rationale behind it. Then, we present an experiment to compare forgetfulness of test-time training under various adversarial scenarios. Last, we discuss the defense tactics and directions for future research.

\section{Methods}

Let $f$ denote a test-time training classifier and $D$ is data $f$ has already seen $D = (x_1, ..., x_n)$. The objective of a lethean attack is to find a sequence $S = (x^*_1, ..., x^*_t)$ such that after $f$ adapts to $S$, its accuracy on samples from $D$ is not better than chance.

The theoretical basis for test-time training lies in the following theorem:
\begin{theorem}[{{\cite{sun2019testtime}}}]
Let $l_m(x, y; \theta)$ denote the main task loss on test instance $x, y$ with parameters $\theta$, and $l_s(x; \theta)$ the self-supervised task loss that only depends on $x$. Assume that for all $x, y, l_m(x, y; \theta)$
is differentiable, convex and $\beta$-smooth in $\theta$, and both $||\nabla l_m(x, y; \theta)|| , ||\nabla l_s(x, \theta)|| \leq G$ for all $\theta$. With a fixed learning rate $\eta = \frac{\epsilon}{\beta G}$, for every $x, y$ such that
\begin{equation}
    \langle \nabla l_m(x, y; \theta), \nabla l_s(x; \theta)\rangle > \epsilon
\end{equation}
We have
\begin{equation}
    l_m(x, y; \theta) > l_m(x, y; \theta(x))
\end{equation}
where $\theta(x) = \theta - \eta \nabla l_s(x; \theta)$ i.e - test-time training with one step of gradient descent.
\end{theorem}

Meaning, as long as the losses for the main and auxiliary tasks are positively correlated, test-time training is expected to perform well.

An adversary could take advantage of the above assumption by crafting samples for which the main and auxiliary losses have a negative correlation. Formally, find $x^*$ for which:
\begin{equation}
    \langle \nabla l_m(x^*, y; \theta), \nabla l_s(x^*; \theta)\rangle < 0
\end{equation}

Notice the difference between standard adversarial examples \citep{szegedy2013intriguing, goodfellow2014explaining} and lethean attacks. We do not require $x$ to be indistinguishable or even similar to a real sample point. The adversary objective is to poison the model for future instances, rather than alter the classification of a present instance.

For certain loss functions and data distributions we could compute $x$ for (3). Instead, we apply a trick. Remember that $f$ had already seen and therefore trained on samples from $D$. Thus, we can expect that the main and auxiliary gradient losses are positively correlated.
\begin{equation}
    \mathbb{E}_{x \in D}\langle \nabla l_s(x, y; \theta), \nabla l_m(x; \theta)\rangle > 0 
\end{equation}
Consequently, we can search for $x^*$ for which:
\begin{equation}
    \mathbb{E}_{x \in D}\langle \nabla l_m(x, y; \theta), \nabla l_m(x^*; \theta)\rangle > 0
\end{equation}
\begin{equation}
    \mathbb{E}_{x \in D}\langle \nabla l_s(x, y; \theta), \nabla l_s(x^*; \theta)\rangle < 0
\end{equation}
If (4), (5), (6) are very strongly correlated then (3) is implied.

In simple words, we try to find $x^*$ which on one hand, positively correlate with the main historical gradient loss and on the other hand, negatively correlate with the auxiliary historical gradient loss. Note that due to symmetry, samples that have negative correlation with the main historical task and positive correlation with the historical sub-task would also work, although we argue that in reality these would be harder to conjure. On the next section I present a practical example.

\section[Experiments]{Experiments\footnote{The code used to conduct experiments is available at \texttt{https://github.com/eyalperry88/lethean}}}

\subsection{Test-Time Training Implementation}
Based on the code released by \cite{sun2019testtime}, we trained from scratch a test-time training network based on the ResNet18 architecture \citep{he2016identity}. The model has two heads which correspond to the classification task and the self-supervised task. The auxiliary task is prediction of rotation by a fixed angle (0$^\circ$, 90$^\circ$, 180$^\circ$, 270$^\circ$). ResNet18 has four "groups", and the split point of the shared parameters between the two tasks is right after the first three. Similarly to the original implementation, we used Group Normalization \citep{wu2018group} to prevent inaccuracies for small batches. The model was trained on the CIFAR-10 dataset \citep{krizhevsky2009learning}, including data augmentations. Optimization was done using stochastic gradient descent (SGD) with momentum and weight decay; learning rate starts at 0.1 and is dropped by 10\% every 50 epochs. Training was done on Nvidia GTX 1080 Ti with batch size 128, for 137 epochs. The final test accuracy of the trained model is 90.2\%.

At test-time, new instances are evaluated sequentially. Each test instance $x$ is rotated in all four angles, for which the auxiliary loss is computed. The model parameters are updated according the auxiliary loss with a single step of learning rate 0.001. 

\subsection{Lethean Attack in Practice}

To perform a lethean attack, we need to craft a sample that is (1) positively correlated with classification gradient loss for CIFAR-10 training data and (2) negatively correlated with rotation gradient loss for the same data. To achieve (1), we pick a sample from CIFAR-10 training data. For (2), we rotate the image by 90$^\circ$, 180$^\circ$ or 270$^\circ$. This simple change is expected to cause a negative correlation between the gradient loss of our adversarial one and the original image that was not rotated. 

The experiment procedure is as follows: at each time step, pick a random sample from the training set, rotate and feed it to the online network. Save the adapted network for future time steps. Every 50 time steps, we evaluate the performance of the network (without adaptation) on the CIFAR-10 test set. Repeat until we reach coin-flip accuracy (\textasciitilde10\% for CIFAR-10).

It is well known that online learning algorithms are naturally prone to forgetfulness, without the need for data poisoning. Moreover, it could be that existing adversarial methods, such as the Fast Sign Gradient Method (FGSM) \citep{goodfellow2014explaining}, would disrupt the model just as bad. To prove the effectiveness of lethean attack, we run the exact same procedure using three other methods: (1) random pixel images, (2) distribution shifts and (3) FGSM attacks. For (1), we generate images where each pixel value is drawn from a normal distribution with the same mean and variance of pixels in the training set. For (2), we follow the evaluation framework used by \cite{sun2019testtime} by picking a random sample from CIFAR-10-C \citep{hendrycks2019benchmarking}, a dataset of noisy and corrupted images based on CIFAR-10. The dataset contains 15 types of noise and perturbations, each with five levels of intensity. We evaluated three types of noise (Gaussian, Shot, Impulse) with the highest intensity level (5). Since all noise types gave extremely similar results, we present only the effect of Gaussian noise. For (3), we pick a random sample from the training set and run the network once (with no adaption). We get the sign of the gradient for each pixel in the the image and use it to perturbate ($\epsilon$ = 0.2) the image, which is then fed to the online model.

\begin{figure}[htp]
  \centering
  \includegraphics[width=10cm]{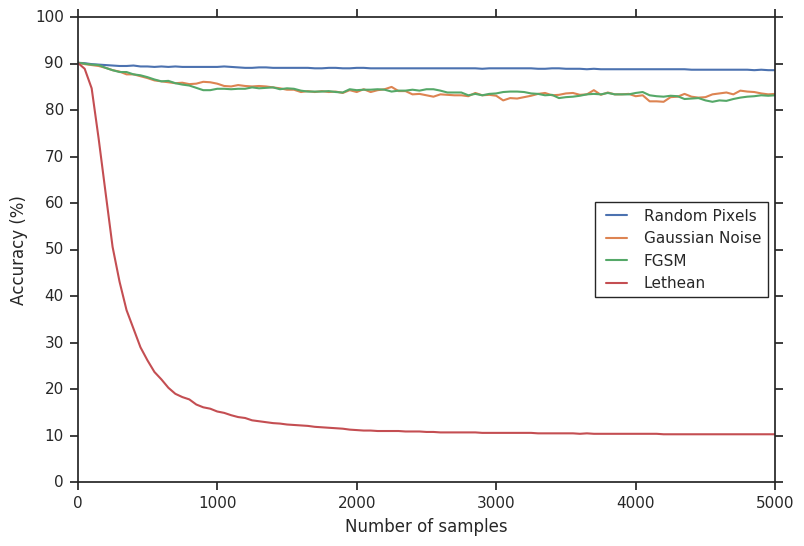}
  \caption{Test-time training forgetting under adversarial scenarios.}
  \label{fig:forgetness}
\end{figure}

The results are presented in Figure 1. The only method to induce complete forgetting is the lethean attack. Only a few dozens of examples are needed to heavily disrupt test performance, and after 1000 examples the model is back to coin-flip accuracy. Random pixels do not affect the network, while both distribution shifts and FGSM attacks do cause some forgetting, but at a significantly slower rate and smaller magnitude.

\section{Discussion}

One defense tactic for test-time training against lethean attacks could be a different auxiliary loss function which isn't as susceptible to malicious examples. A more robust approach could be regularization that controls for the correlations between new gradient updates and historical ones. Hence, limit the learning step to be always somewhat correlated with previous learning steps. We cannot apply this method at the beginning of training, but once a model reaches high performance, activating correlation regularization could prevent "untraining" the model back to coin-flip accuracy. A drawback of this scheme is adaptation to abrupt distribution shifts, therefore it is more fitting for real-life scenarios where distribution shifts are smooth and have a small Lipschitz constant. It is worth noting that in \cite{sun2019testtime} implementation of test-time training, there exists a hyper-parameter \textit{threshold confidence}. This parameter controls which samples are being trained on, such that samples which the model is highly confident about, do not need to alter the model. From our preliminary experimentation, this parameter slows down but does not prevent a lethean attack.

Since catastrophic forgetting is an inherent feature of current online neural networks, test-time training could benefit from the field of continual learning \citep{li2017learning, lopez2017gradient, kirkpatrick2017overcoming} which aims to prevent forgetting in general, not necessarily under an adversarial setting. Understanding forgetting is a major milestone in neural networks research, as it is one of the most significant dissimilarities between artificial and biological neural networks. Last, the concept of lethean attacks, a.k.a memory erasure in human minds has been the subject of countless Hollywood movies\footnote{A few personal recommendations: \textit{Eternal Sunshine of the Spotless Mind, Inception, Men in Black}} and in reality could potentially restore the lives of millions of PTSD patients. A review of the controversial field of memory reconsolidation \citep{besnard2012reconsolidation} and its applications is beyond the scope of this paper, interesting as they are.

\section{Conclusion}

Breaking the boundary between training and test-time enables generalization across distribution shifts, and possibly other interesting applications such as hyper-personalization. In this work we examined the tradeoff which emerges from model modification in an online setting. A malicious agent with limited knowledge could render a test-time trained model completely useless. We do not consider this work as a warning against adaptive online learning paradigms, on the contrary, we would like to aid the field's development and leverage lethean attacks as a framework to analyze an online model robustness.

\bibliography{biblio}

\end{document}